\documentclass[english,aps,superscriptaddress,manuscript]{revtex4-1}
\usepackage[colorlinks=true,citecolor=blue,linkcolor=blue]{hyperref}
\usepackage{graphicx}
\usepackage{color}
\usepackage{dcolumn}
\usepackage{bm}
\usepackage{mathrsfs}
\usepackage{amsmath}
\usepackage{amssymb}
\usepackage{amsthm}
\usepackage{amsfonts}
\usepackage{enumerate}
\usepackage{latexsym}
\usepackage{hyperref}
\usepackage{centernot}
\usepackage{multirow}

\begin{document}

\title{Robust boundary flow in chiral active fluid} %

\begin{abstract}
We perform experiments on an active chiral fluid system of self-spinning
rotors in confining boundary. Along the boundary, actively rotating
rotors collectively drives a unidirectional material flow. We systematically
vary rotor density and boundary shape; boundary flow robustly emerges
under all conditions. Flow strength initially increases then decreases
with rotor density (quantified by area fraction $\phi$); peak strength
appears around a density $\phi=0.65$. Boundary curvature plays an
important role: flow near a concave boundary is stronger than that
near a flat or convex boundary in the same confinements. Our experimental
results in all cases can be reproduced by a continuum theory with
single free fitting parameter, which describes the frictional property
of the boundary. Our results support the idea that boundary flow in
active chiral fluid is topologically protected; such robust flow can
be used to develop materials with novel functions.
\end{abstract}

\date{\today }
\author{Xiang Yang$^\ddag$}
\email{yangxiang816@sjtu.edu.cn}
\affiliation{School of Physics and Astronomy and Institute of Natural Sciences,
Shanghai Jiao Tong University, Shanghai, China}
\affiliation{School of Physcial Science and Technology, Soochow
University, Suzhou, China}

\author{Chenyang Ren$^\ddag$}
\affiliation{School of Physics and Astronomy and Institute of Natural Sciences,
Shanghai Jiao Tong University, Shanghai, China}

\author{Kangjun Cheng}
\affiliation{Zhiyuan College, Shanghai Jiao Tong University, Shanghai, China}

\author{H. P. Zhang}
\email{hepeng\_zhang@sjtu.edu.cn}
\affiliation{School of Physics and Astronomy and Institute of Natural Sciences,
Shanghai Jiao Tong University, Shanghai, China}
\affiliation{Collaborative Innovation Center of Advanced Microstructures, Nanjing, China}

\maketitle




\section{Introduction}

Active matter is composed of constituent units individually powered
by internal or external energy sources. In majority of current studies,
local energy injection drives constituent unit's linear motion \citep{Marchetti2013}.
In these systems, a wide range of phenomena has been reported, including
emergent collective motion \citep{Bricard2013,Kumar2014}, pattern
formation \citep{Riedel2005,Cates2010,Farrell2012,Bricard2015} and
phase segregation without attraction \citep{Fily2012,Redner2013,Stenhammar2013}.
Besides linear motion, local energy injection can also cause constituent
unit to actively rotate. Biological examples of such chiral active
matter include rotating bacteria \citep{Petroff,Chen2015}, circling
bacteria \citep{Diluzio2005,Lauga2006,Leonardo2011} and sperm cells
\citep{Friedrich2007,Riedel2005} near surfaces, and magnetotactic
bacteria in rotating fields \citep{Erglis2007,Cebers2011}. Artificial
chiral active systems have also been developed, such as colloids \citep{Yan2015,Yan2015c,Maggi2015,Kokot2017,Xie2019,Aubret2018,Soni2019},
millimeter-scale magnets\citep{Grzybowski2000,Grzybowski2002} and
rotating granular particles \citep{Tsai2005,Scholz2018,Farhadi2018,Workamp2018}.
Multiple numerical and theoretical studies on chiral active fluid have been carried
out \citep{Lenz2003,Tsai2005,Furthauer2013,Nguyen2014a,Spellings2015,Goto2015,Aragones2016,Yeo2015,Reichhardt2019}.

Interacting active rotors can form a range of collective phenomena.
One of such phenomena is unidirectional material flow localized at
rotor/solid \citep{Tsai2005,Zuiden2016}, rotor/liquid \citep{Soni2019}
and rotor phase boundaries \citep{Nguyen2014a,Scholz2018}. A continuum
theory was developed to reproduce boundary flow in a driven granular system
in a circular confinement \citep{Tsai2005}. Later, the same theory
was compared with numerical data of confined rotors \citep{Zuiden2016}.
Recently, Dasbiswas, Mandadapu, and Vaikuntanathan \citep{Dasbiswas2018}
studied topological properties of the continuum theory; they showed
that the emergence of the boundary flow in active chiral fluid can
be understood as an example of topological protection at boundary
\citep{Dasbiswas2018} and that the boundary flow is insensitive to
boundary interactions and highly resistant to perturbations. Similar
robustness has been extensively studied in many topologically nontrivial
systems, such as mechanical lattice \citep{Kane2013,Paulose2015,Lubensky2015},
electronic \citep{Hasan2010} and photonic \citep{Haldane2008} systems.

Here, we investigate boundary flow of individually-driven, rotating
particles in confining boundaries by experiment and theory. Our experiments
show that boundary flow robustly emerges in all cases of various rotor
densities and boundary shapes. To facilitate the comparison between
experiments and theory, we use experimental observations to simplify
the continuum theory \citep{Tsai2005} and carry out independent
experiments to identify model parameters. Eventually, our
experimental results in all cases can be reproduced by the continuum
theory with single free fitting parameter, which describes the frictional
property of the boundary.

\begin{figure}[ht]
\includegraphics[width=8cm]{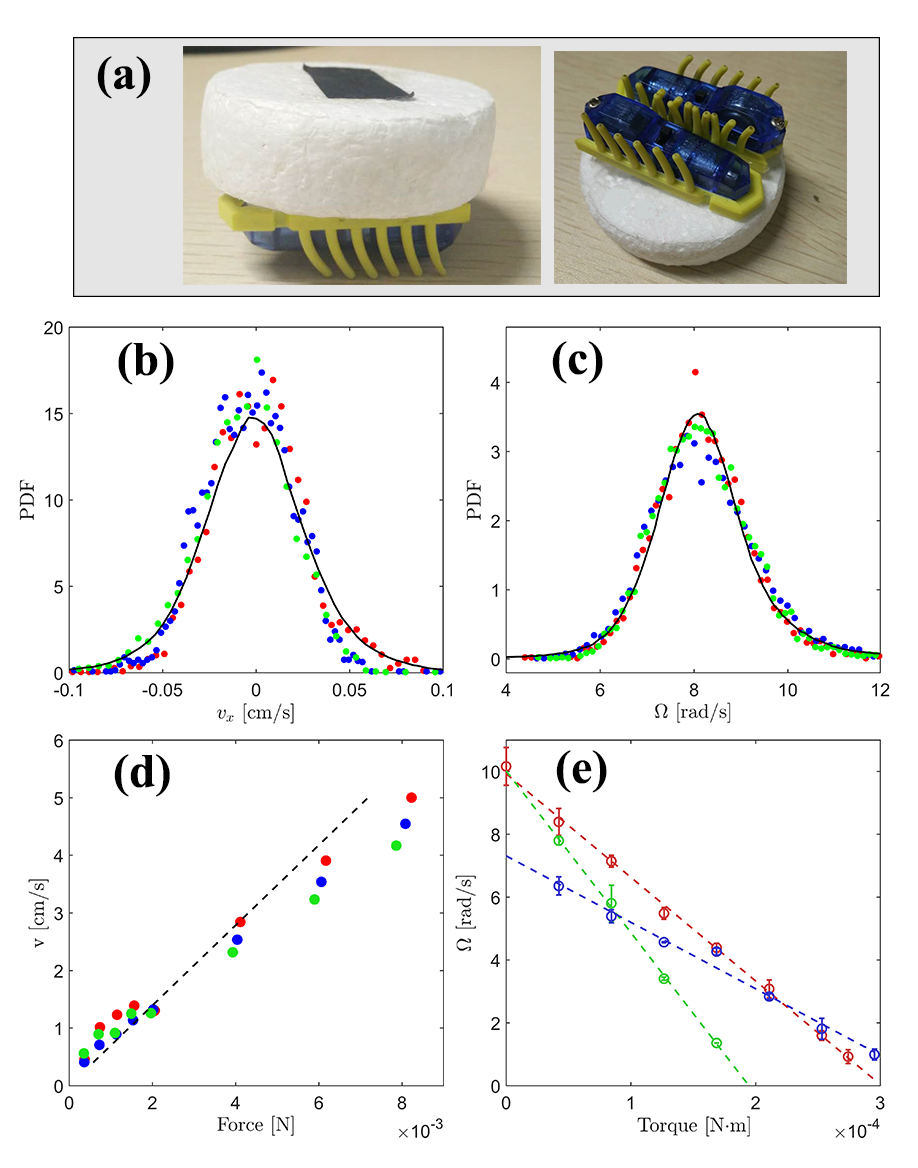}
\centering
\caption{Construction and properties of single rotor. (a) two images showing
a rotor made of two Hexbug robots and a foam disk. (b-c) Probability
distribution functions of  linear velocity and spin rate. Lines and symbols represent
average results over 40 rotors and data from individual rotors, respectively. (d-e) Rotor
responses to external force and torque. Lines are linear fits to experimental
data (symbols). }
\end{figure}

\section{Experimental methods and results}

Our rotor ($m=15.3\pm0.2$g in mass) is driven by two Hexbug robots.
Each Hexbug, 4.3 cm long and 1.2 cm wide, houses a 1.5V button cell
battery that drives a vibration motor; we use fresh batteries in each
new experiment and run experiments for less than 20 minutes to prevent
battery power degrading. Hexbug body is supported by twelve flexible
legs that all bends slightly backwards. When turned on, the vibration
motor sets Hexbug into forward hopping motion on a solid (PMMA) substrate
\citep{Dauchot2019,Li2013}. As shown in Fig. 1(a), two Hexbug robots
in a rotor are glued to a foam disk ( radius $a=2.5$cm ) in opposite
directions; they can generate a torque that spins the rotor with a
spin rate about $\varOmega_{0}\approx8.4$ rad/s, cf. Fig. 1(c). Rotors
and the substrate are carefully balanced so that translational motion
of an isolated rotor is suppressed, cf. Fig. 1(b). Our rotors respond
linearly to external force and torque, as shown in Fig. 1(d-e); detailed
description of these experiments can be found in \textbf{supplementary information}.

To observe localized boundary flow, we confine rotors with solid boundaries
which are precisely machined by a laser-cutter and covered with smooth
tapes to reduce friction, cf. Fig. 2(a). Different numbers of rotors
are used to vary density. Rotor motion is recorded by a digital camera
at 30 frames per second; we use standard particle tracking method
to measure rotor translation and rotation from recorded videos. Experimental
results obtained in both axisymmetric and non-axisymmetric confinements
are discussed in detail below.

\begin{figure*}[ht]
\includegraphics[width=17cm]{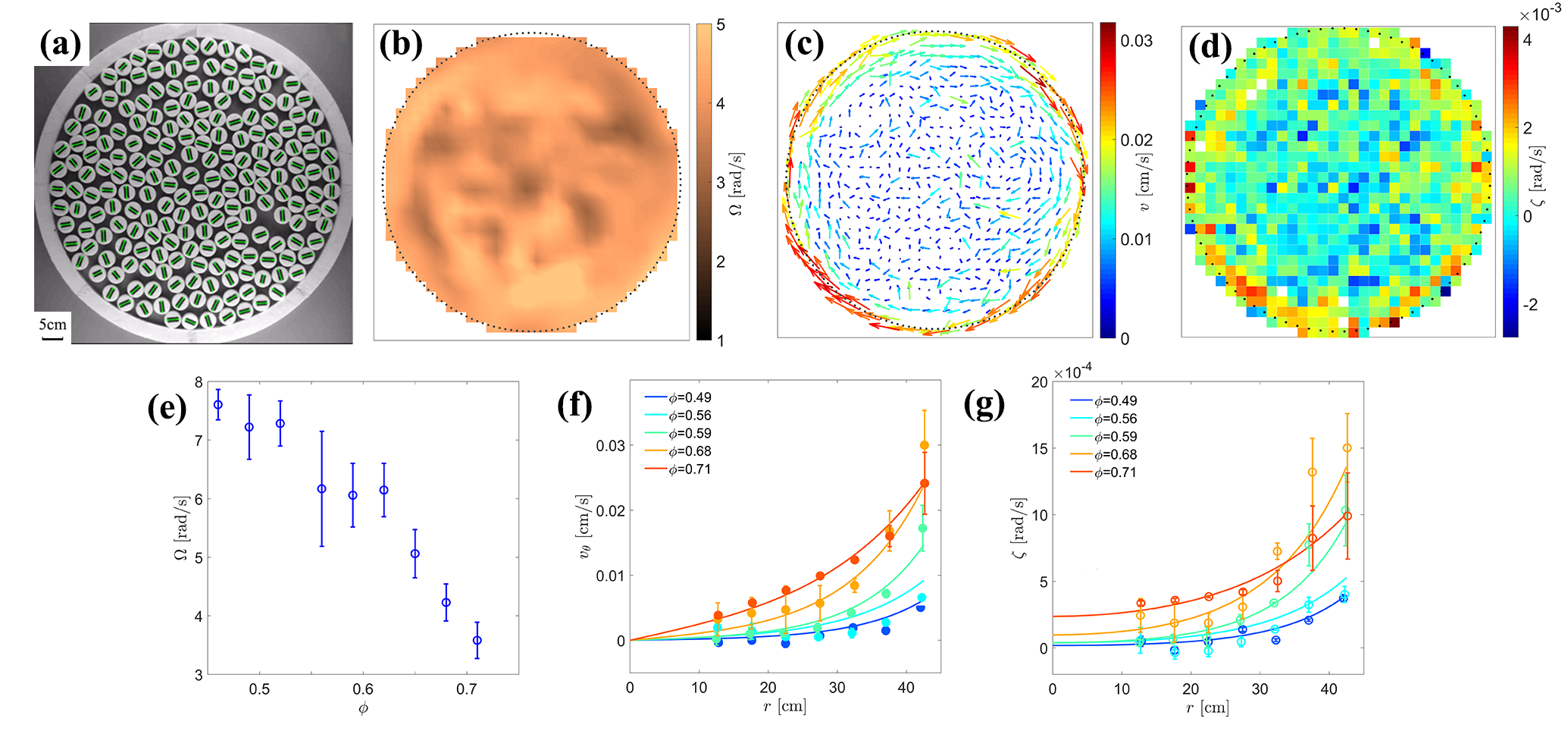}
\centering
\caption{Results in a circular confinement. (a) Snapshot from an experiment
with $\phi=$0.68. Black markers on rotors are used to facilitate
particle tracking. Coarsen-grained fields ($\varOmega(\vec{r})$,
$\vec{v}(\vec{r})$ and $\zeta(\vec{r})$) from the same experiment
are shown in (b-d); Dotted line marks the effective boundary of the
space accessible to the rotor center. (e) Density dependence of averaged
spin rate $\Omega$. (f-g) Radial profiles of $v_{\theta}(r)$ and
$\zeta(\vec{r})$. symbols and lines represent experimental and numerical
results, respectively. }
\end{figure*}

\subsection{Axisymmetric boundary}

We start from a circular boundary with a radius $R_{c}=45$ cm, cf.
Fig. 2(a). Five different numbers of rotors are used: $N=$ 160,180,190,220
and 230; the corresponding area fraction $\phi\equiv\frac{N\pi a^{2}}{\pi R_{c}^{2}}$
are 0.49, 0.56, 0.59, 0.68 and 0.71, respectively. Typical system
dynamics can be seen in \textbf{supplementary movie S1}: while spinning
rotors interact with neighbors and boundary, part of their angular
momentum is converted to linear momentum, which is reflected by rotors'
translational motion; rotor translation is most pronounced near the
boundary and is in the clockwise direction.

We measure spin rate $\varOmega$ and velocity $\vec{v}$ of each
rotor and average measured results in 1.5$\times$1.5 c${\textstyle \textrm{m}}^{2}$
bins. As shown in Fig. 2(b) and (e), coarse-grained spin rate $\varOmega(\vec{r})$
has an approximately uniform spatial distribution and its mean value
decreases as the rotor density increases, cf. Fig. 2(e); this is mainly caused by
the frictional slides of neighbors. Coarsen-grained linear velocity
field $\vec{v}(\vec{r})$ is shown in Fig. 2(c); local
angular velocity $\zeta(\vec{r})\equiv\frac{1}{2}\left(\nabla\times\vec{v}(\vec{r})\right)_{z}$
computed and plotted in Fig. 2(d). By averaging data in concentric
annuli between $r-a$ and $r+a$, we can get radial profiles of $v_{\theta}(r)$
and $\zeta(r)$. Data in Fig. 2(f-g) show that localized boundary
flow emerges under all density conditions with different strength.

We add an inner boundary ($15$ cm in radius) to the system; this
makes a ring-shaped confinement, as shown in Fig. 3(a).
As in the case of circular boundary, coarsen-grained fields, $\vec{v}(\vec{r})$
and $\zeta(\vec{r})$, and their radial profiles are measured and
plotted in Fig. 3. From these data, we see that, in addition to the
clockwise flow along the outer boundary, a counter-clockwise flow
emerges near the inner boundary, which is weaker and can most clearly
seen from $v_{\theta}(r)$ profiles in Fig. 3(c).

\subsection{Non-axisymmetric boundary}

We further investigate two cases of non-axisymmetric boundary: capsule-shaped
and U-shaped confinements as shown in Fig. 4(a) and
(d). In both cases, curvature changes along the boundary and affects
the strength of boundary flow. To quantify this point, linear velocity
and local angular velocity at representative points, square and
circular symbols in (a) and (d), are computed at six rotor densities.
Data in Fig. 4(b-c) show boundary flow near concave boundary (red
symbols) is stronger than that (blue symbols) near a flat region.
In the case of U-shaped confinement, cf. Fig. 4 (d-f) and \textbf{supplementary movie S2},
concave boundary(red symbols) generates stronger flow than convex boundary (blue symbols).
We also discover that the flow velocity peaks near the density $\phi=0.65$
in both cases, as shown in Fig. 4(b) and (e).

\begin{figure}[h]
\includegraphics[width=9cm]{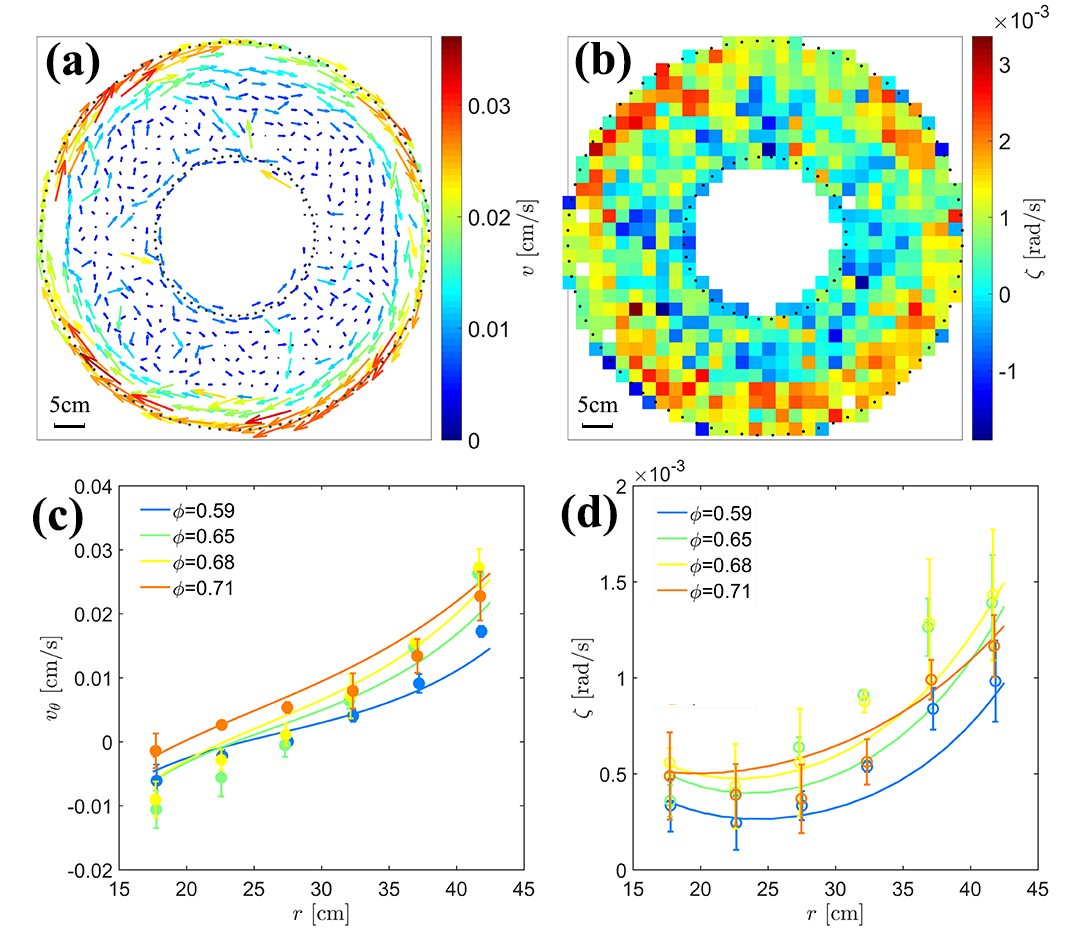}

\caption{Results in a ring-shaped confinement. (a) velocity field and (b) local
collective angular velocity field measured in experiments with $\phi=$0.68.
Radial profiles of $v_{\theta}(r)$ and $\zeta(\vec{r})$ are shown
in (c-d); symbols and lines represent experimental and numerical results,
respectively.}
\end{figure}

\begin{figure*}[t]
\includegraphics[width=14cm]{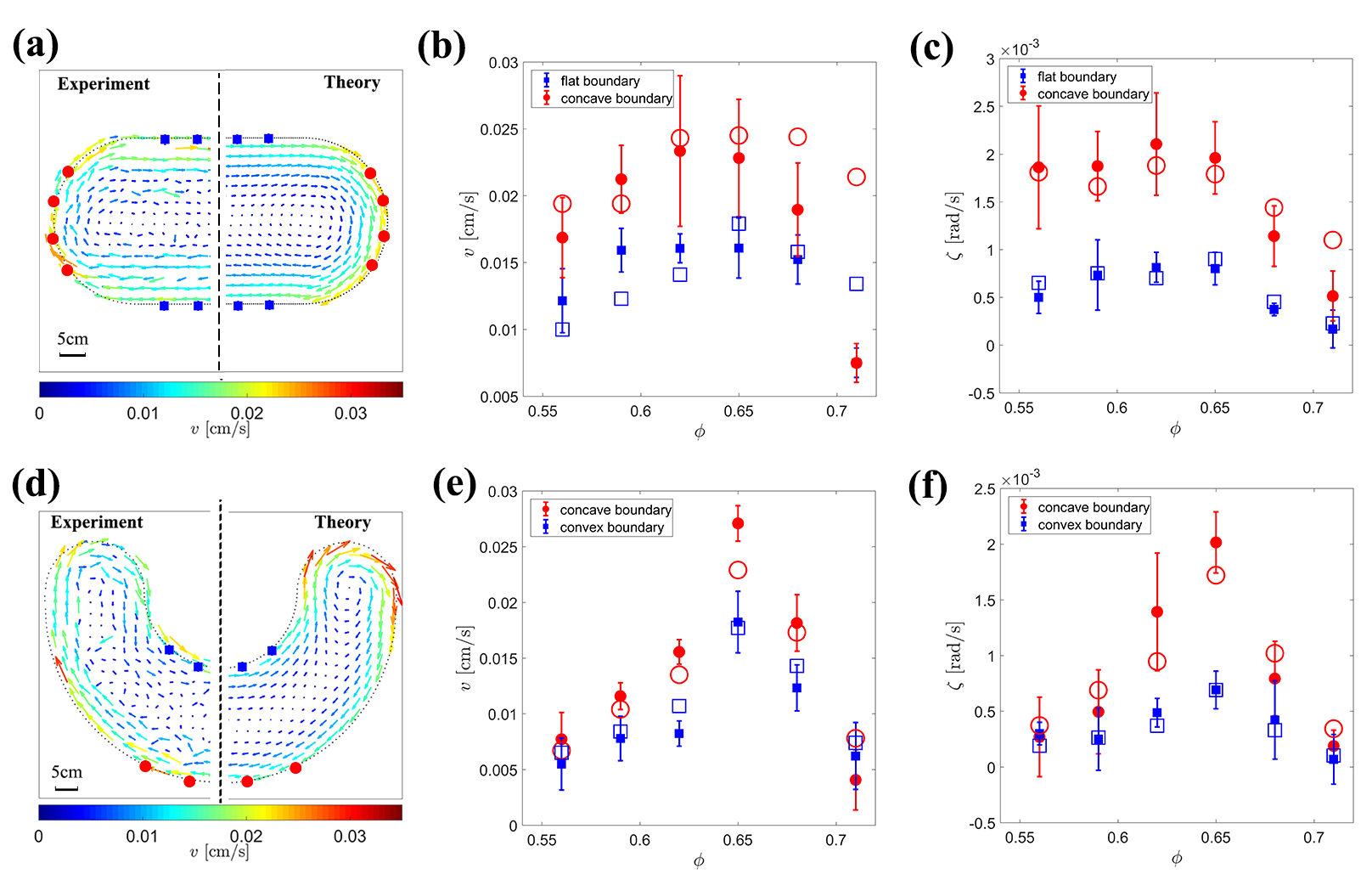}
\centering
\caption{Results in two non-axisymmetric cases: (a-c) capsule-shaped and (d-f)
U-shaped confinements. Effective boundary of rotor centers in two cases is represented by dotted line in (a) and (d). Experimentally measured velocity fields with
$\phi=$0.68 are shown in the left half of (a) and (d) with corresponding
theoretical prediction on the right. Averaged velocity and local angular
velocity at representative points, marked by symbols in (a) and (d),
are computed at different densities and plotted in (b-c) and (e-f);
experimental and theoretical results are represented by filled and
empty symbols, respectively.}
\end{figure*}

\section{Theoretical analysis of experimental results}

\subsection{Continuum theory}

To understand experimental results in Figs. 1-4, we use a continuum
theory developed by Tsai and coauthors \citep{Tsai2005}. The theory describes the conservation
laws of the following hydrodynamic variables: the mass density $\rho(\vec{r},t)$,
the momentum density $\rho\vec{v}(\vec{r},t)$ and the angular momentum
density $I\varOmega(\vec{r},t)$, where $I$ is the moment of inertial
density. The first continuum equation describes mass conservation:

\begin{equation}
\frac{\partial\rho}{\partial t}+\nabla\cdot(\rho\vec{v})=0,\label{eq:continuum}
\end{equation}
where the mass density $\rho$ is proportional to the area fraction
of rotors $\phi$: $\rho=\frac{m}{\pi a^{2}}\phi$. Rotor density
in our experiments is spatially homogeneous, cf. Fig. 2(a) and Fig.
S2, which allows us simplify Eq. (\ref{eq:continuum}) as :
\begin{equation}
\nabla\cdot\vec{v}=0.\label{eq:Sim_v1}
\end{equation}

The angular momentum of rotors is conserved:

\begin{equation}
ID_{t}\varOmega=D_{\varOmega}\nabla^{2}\varOmega-\Gamma^{\varOmega}\varOmega-\Gamma(\varOmega-\zeta)+\tau,\label{eq:angular momentum 1}
\end{equation}
where $D_{t}\equiv\partial_{t}+\vec{v}\cdot\nabla$ is convective
derivative, $D_{\varOmega}$ is the angular momentum diffusion constant,
$\Gamma^{\varOmega}$ is the angular friction coefficient due to rotor-substrate
interaction, $\Gamma$ is spin-velocity coupling constant, and $\tau$
stands for driving torque density field experienced by the rotors.
We can simplify Eq. (\ref{eq:angular momentum 1}) as

\[
\tau=\Gamma^{\varOmega}\varOmega+\Gamma\varOmega
\]
by the following experimental observations: 1) our system in steady
state; 2) homogeneous angular momentum field $\varOmega$ (Fig. 2b);
3) local angular velocity $\zeta$ is much less than spin rate $\varOmega$
(Fig. 2g and 3d). Under low density condition, isolated rotors experiences
little coupling to others, i.e. $\Gamma=0$, we have spin rate
for isolated rotors:
\[
\varOmega_{0}=\frac{\tau}{\Gamma^{\varOmega}}.
\]
Combining two equations above, we have the following relation:
\begin{equation}
\frac{\varOmega}{\varOmega_{0}}=\frac{\Gamma^{\varOmega}/\varGamma}{\Gamma^{\varOmega}/\varGamma+1}.\label{eq:Sim_Om}
\end{equation}

Momentum conservation requires:

\begin{equation}
\rho D_{t}\vec{v}=-\nabla p+\eta\nabla^{2}\vec{v}-\Gamma^{v}\vec{v}+\frac{\Gamma}{2}\epsilon\nabla(\varOmega-\zeta),\label{eq:momentum 1}
\end{equation}
where $\eta$ is the shear viscosity, $\Gamma^{v}$ is the linear
friction coefficient, and $\epsilon$ is 2D antisymmetric symbol.
The odd viscosity has been ignored in our system for quite large damping
coefficient $\Gamma^{v}$ \citep{Banerjee2017}. With a steady-state
assumption, we take the curl of Eq. (\ref{eq:momentum 1}):

\begin{equation}
\left((4\eta+\Gamma)\nabla^{2}-4\Gamma^{v}\right)\zeta-\Gamma\nabla^{2}\varOmega=0.
\end{equation}
The above equation can be further simplified by assuming a homogeneous
angular momentum field $\varOmega$ and weak coupling $\Gamma\ll\eta$
(see Fig. 5b); we end up with the following equation:

\begin{equation}
\left(\nabla^{2}-\frac{\Gamma^{v}}{\eta}\right)(\nabla\times\vec{v})_{z}=0.\label{eq:Sim_v2}
\end{equation}

\subsection{Boundary conditions}

Eq. (\ref{eq:Sim_v1}) and Eq. (\ref{eq:Sim_v2}) can be solved with
proper boundary conditions. The boundary is characterized by a local
outward normal vector, $\hat{r}$, and a tangential direction, $\hat{\theta}$;
the local radius curvature is denoted as $R$ with the convention
that a concave boundary has a positive radius of curvature. A rigid
wall requires the radial velocity component to be zero:
\begin{equation}
v_{r,B}=0,\label{eq:BC1}
\end{equation}
where the subscript ``$B$'' stands for the boundary of the occupied
region for rotor centers, as shown by the dotted line in Fig. 2(b-d)
for circular boundary. The second boundary condition arises from the
fact that rotors also experience a frictional force from the boundary;
this leads to a tangential-radial component of the stress tensor:

\begin{equation}
\sigma_{\theta r,B}=f_{B},
\end{equation}
where $f_{B}$ is an effective boundary friction on unit length. We
assume that $f_{B}$ is proportional to the shear stress from the
spin-velocity coupling:
\begin{equation}
f_{B}=-k\Gamma\varOmega,\label{eq:boundary friction}
\end{equation}
As the friction of rotor-rotor and rotor-boundary
have similar dependence on $\phi$, the proportion factor $k$ in Eq. \ref{eq:boundary friction} is treated as a constant in a given experiment. We express $\sigma_{\theta r}$ in velocity components,
combine Eqs. (\ref{eq:BC1}-\ref{eq:boundary friction})
and obtain the following boundary condition (See \textbf{supplementary information} for detailed discussions):

\begin{equation}
\left(\zeta-\frac{v_{\theta}}{R}\right)_{B}=\frac{\varGamma}{4\eta}(1-2k)\varOmega_{B}.\label{eq:BC2}
\end{equation}

\subsection{Determination of model parameters}

Fig. 1(d) and (e) shows linear responses of isolated rotors to external
force and torque. By measuring slopes of data in these plots, we extracted
linear and angular frictional coefficient for isolated rotors: $\gamma^{v}=0.14$
kg/s and $\gamma^{\varOmega}=0.32$ kg $\textrm{cm}{}^{2}$/s. These
two quantifies are related to $\Gamma^{v}$ and $\Gamma^{\varOmega}$
as : $\Gamma^{v}=\rho\gamma^{v}=\frac{m}{\pi a^{2}}\phi\gamma^{v}$
and $\Gamma^{\varOmega}=\rho\gamma^{\varOmega}=\frac{m}{\pi a^{2}}\phi\gamma^{\varOmega}$;
this leads to $\Gamma^{\varOmega}/\Gamma^{v}=0.44$ cm\textbf{$^{-2}$}
for all densities.

Eq. (\ref{eq:Sim_Om}) relates spin rate $\Omega$ to the ratio of
angular frictional coefficient $\Gamma^{\varOmega}$ to coupling constant
$\Gamma$. Our experiments show that spin rate $\Omega$ decreases
with increasing rotor density, cf. Fig. 2(e). From such data, we can
use Eq. (\ref{eq:Sim_Om}) to measure the ratio $\Gamma^{\varOmega}/\Gamma$
in different confinements and different densities. Results are plotted
in Fig. 5(a), showing a monotonic decrease with area fraction $\phi$;
collapse of all data on a single curve demonstrates that this ratio
depends weakly on boundary shape and is a bulk property of the system.

We can estimate the ratio $\varGamma/\eta$ from stress boundary condition,
Eq. \ref{eq:BC2}, by rewriting the equation as
\[
\frac{\varGamma}{\eta}=\frac{4\left(\zeta-v_{\theta}/R\right)_{B}}{(1-2k)\varOmega_{B}}.
\]
Quantities in the above equation, $\zeta,v_{\theta},\text{and }\Omega$
at boundary, can be measured directly from experiments. Therefore,
for any given proportion constant $k$, we can compute $\frac{\varGamma}{\eta}$
along the boundary then average computed values, which depends weakly
on local curvature. Averaged results for $\frac{\varGamma}{\eta}$
obtained with $k=0.4$ are plotted in Fig. 5(b). $\frac{\varGamma}{\eta}$
results from different confinements approximately collapse onto a
single curve and show a peak around density $\phi=0.65$, where peak
boundary flow in Fig. 4(b) and (e) appears. Fig. 5(b) shows that spin-velocity
coupling is weak in our system, with a coupling constant $\Gamma$
two-order magnitude smaller than the shear viscosity $\eta$.

\subsection{Comparison between theoretical and experimental results}

With the process in the above section, we can estimate three parameter
ratios in every experiment. From these ratios, the only parameter
in Eq. (\ref{eq:Sim_v2}) can be determined:
\begin{equation}
\frac{\Gamma^{v}}{\eta}=\left(\frac{\Gamma^{v}}{\Gamma^{\varOmega}}\right)\left(\frac{\Gamma^{\varOmega}}{\varGamma}\right)\left(\frac{\varGamma}{\eta}\right).\label{eq:lamd}
\end{equation}
Because spin rate $\Omega$ is spatially homogeneous, we set its boundary
value in Eq. (\ref{eq:BC2}) $\Omega_{B}$ as the measured spin rate
in bulk. The proportion constant $k$ is treated as an adjusting parameter.
For a given $k$ value, we use a finite-element package (COMSOL) to
solve Eq. (\ref{eq:Sim_v1}) and Eq. (\ref{eq:Sim_v2}) with boundary
conditions Eq. (\ref{eq:BC1}) and Eq. (\ref{eq:BC2}) for steady
flow $\vec{v}(\vec{r},t)$ in all experiments. In a typical calculation,
more than 10,000 finite elements are used to ensure convergence. Detailed numerical
results for a capsule-shaped confinement can be found in Fig. S4.

We compare theoretical results to experiments and find that $k=0.4$
yields the best overall agreement. All theoretical results in Figs.
2-4 are computed with $k=0.4$. In the case of circular boundary,
theoretical solutions correctly capture the spatial lengthscale and
density dependence of the boundary flow, as shown in Fig. 2. In ring
geometry, Fig. 3, continuum theory predicts the reversal of flow direction
as one moves from the inner to outer boundary and non-monotonic behavior
in local angular velocity, $\zeta$. In non-axisymmetric cases, main
features of steady flow are well captured in theoretical solutions,
especially how flow depends on local curvature and rotor density.
Effect of local curvature is manifested through the stress boundary
condition, Eq. (\ref{eq:BC2}), see detailed discussion in \textbf{supplementary information}.
Rotor density enters the theory through model parameters shown in
Fig. 5.

We note that Fig. 4 (b) and (e) show some deviation of theoretical
results from experiments at high densities. This is likely caused
by transient jamming of densely packed rotors in experiments, which
are not captured in the current fluid-based continuum theory. Transient
jamming and associated elastic stress can also explain the sharp drop
of boundary flow beyond density $\phi=0.65$, cf. Fig. 4 (b) and (e).

\begin{figure}
\includegraphics[width=9cm]{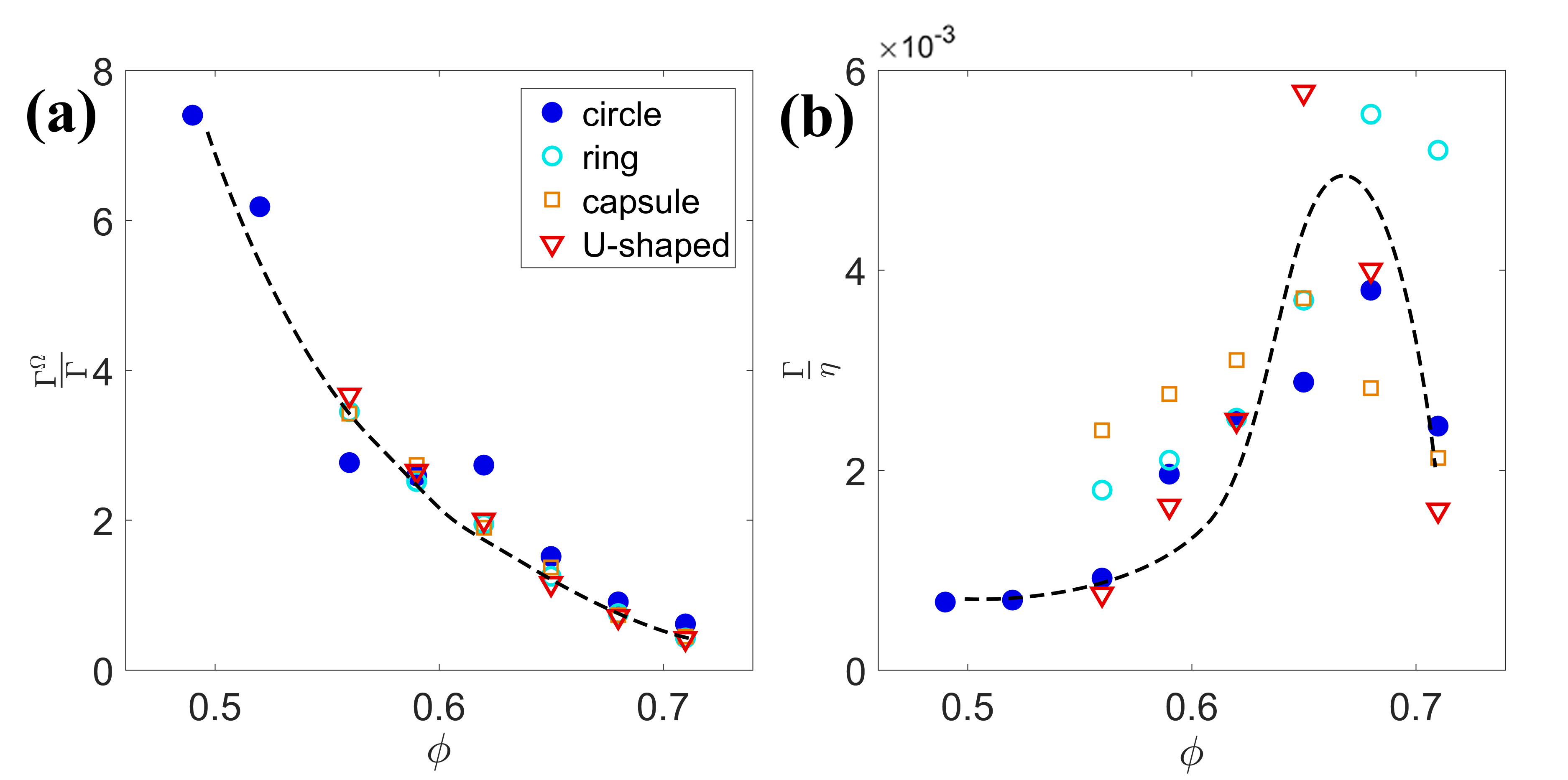}

\caption{Density dependence of two parameter ratios extracted from experiments: (a) $\frac{\Gamma^{\varOmega}}{\varGamma}$
and (b) $\frac{\varGamma}{\eta}$. Data from different confinements
are shown by symbols. Dashed lines are a guide to the eye. }
\end{figure}

\section{Conclusion and discussion}

We have studied collective dynamics of rotors in various confining
boundaries and density conditions. Actively rotating rotors collectively drives a unidirectional
material flow along boundary. Boundary flow robustly emerges in all
experiments with different rotor densities and boundary shapes. We
showed that flow strength initially increases then decreases with
rotor density and peak strength appears around a density $\phi=0.65$.
Boundary curvature plays an important role: flow near a concave boundary
(with a positive radius of curvature) is stronger than that near a
flat or convex boundary in the same confinements. We corroborate experimental
measurements with a theoretical analysis based on a continuum theory,
which is simplified under our experimental conditions; independent
experimental measurements were used to determine transport coefficients
in the theory. We demonstrated that our experimental results in all
cases were quantitatively reproduced by the theory with single free
fitting parameter, which describes the frictional property of the
boundary. Our experimental and theoretical results support the idea
that emergence of robust boundary flow in chiral active matter is
an example of topological protection phenomena in dissipative system
\citep{Dasbiswas2018}. Topological nature of the boundary flow may
allow us to develop new materials with novel and robust functions.

\section*{Acknowledgements}
We acknowledge financial support from National Natural Science Foundation of China Grants (11422427, 11774222 and 11674236) and from the Program for Professor of Special Appointment at Shanghai Institutions of Higher Learning (Grant GZ2016004). We thank the Student Innovation Center at Shanghai Jiao Tong University for support. We thank Xiaqing Shi and Dezhuan Han for useful discussions.
$^\ddag$ X.Y. and C.R. contributed equally to this work.


\bibliography{ActiveFluid}

\end{document}